\documentstyle[12pt]{article}
\begin{document}

{\large Nonlinear Analysis}: {\small Modelling and Control}, Vilnius, IMI,
Vol. {\bf 2}, p. 43-58 (1998)

\vspace{1.5cm}

\begin{center}
{\large {\bf MAPS FOR\ ANALYSIS\ OF\ NONLINEAR\ DYNAMICS }}

\vspace{0.5cm}

{\large B. Kaulakys}

\vspace{0.5cm}

{\it Institute of Theoretical Physics and Astronomy, A. Go\v stauo 12, 2600
Vilnius, Lithuania and Vilnius University, Department of Quantum Electronics
}\vspace{0.5cm}

\parbox{6in}{\small Area preserving maps provide the simplest and most
accurate means to visualize and quantify the behavior of nonlinear systems.
Convenience of the mapping equations of motion for investigation of
transition to chaotic behavior in dynamics of classical atom in microwave
field, transition to nonchaotic behavior in randomly driven
systems and induced quantum dynamics of simple and multilevel systems
is demonstrated.}

\vspace{1cm}
\end{center}

\section{Introduction}

A common method of displaying the dynamics is through a Poincar\'e section.
The Poincar\'e section is a device invented by Henri Poincar\'e as a means
of simplifying phase space diagrams of complicated systems. It is
constructed by viewing the phase space diagram stroboscopically in such a
way that the motion is observed periodically. A dynamical system whose phase
space is tree-dimensional may be converted through the Poincar\'e section to
a two-dimensional mapping. Sometimes such mappings may be further simplified
to one-dimensional. Because of their relative simplicity one- and
two-dimensional maps provide several advantages over the differential
equations. They allow for simple reveal of many characteristics of chaotic
behavior, such as sensitivity to initial conditions, illustration the
mechanisms of bifurcation and so on.

Area preserving maps provide the simplest and most accurate means to
visualize and quantify the behavior of conservative systems. Such maps may
be iterated on even the simple computer. For analysis of the
non-conservative systems sometimes the non-area-preserving maps may be
introduced.

Here we demonstrate examples of investigation of transition to chaotic
behavior in the dynamics of classical atom in microwave field by
area-preserving maps, transition to nonchaotic behavior in randomly driven
systems by the non-area-preserving maps and induced quantum dynamics of
simple and multilevel systems represented by the appropriate quantum maps.

\section{Kepler maps}

Consider dynamics of the classical hydrogenic atom in the monochromatic
field. The direct way of coupling the electromagnetic field to the electron
Hamiltonian is through the ${\bf A\cdot P}$ interaction, where ${\bf A}$ is
the vector potential of the field and ${\bf P}$ is the generalized momentum
of the electron. The Hamiltonian of the hydrogen atom in a linearly
polarized field $F\cos (\omega t+\vartheta )$, with $F$, $\omega $ and $
\vartheta $ being the field strength amplitude, field frequency and phase,
respectively, in atomic units is
$$
H={\frac 12}\left( {\bf P}-{\frac{{\bf F}}\omega }\sin (\omega t+\vartheta
)\right) ^2-{\frac 1r}.\eqno{(2.1)}
$$
Electron energy change due to interaction with the external field follows
from the Hamiltonian equations of motion [1]
$$
\dot E=-{\bf \dot r\cdot F}\cos (\omega t+\vartheta ).\eqno{(2.2)}
$$
Using parametric equations of motion in the Coulomb field we may calculate
the change of the electron's energy in the classical perturbation theory
approximation.

Measuring the time of the field action in the field periods one may
introduce the scale transformation where the scaled field strength and the
scaled energy are $F_s=F/\omega ^{4/3}$ and $E_s=E/\omega ^{2/3}$,
respectively. However, it is convenient [2, 3] to introduce the positive
scaled energy $\varepsilon =-2E_s$ and the relative field strength $
F_0=Fn_0^4=F_s/\varepsilon _0^2$, with $n_0$ being the initial effective
principle quantum number, $n_0=\left( -2E_0\right) ^{-1/2}$. The threshold
values of the relative field strength $F_0$ for the ionization onset depends
weaker upon the initial effective principle quantum number $n_0$ and the
relative frequency of the field $s_0=\omega n_0^3$ than the scaled field
strength $F_s$.

We restrict our subsequent consideration to the one-dimensional model, which
corresponds to the states of low orbital quantum numbers $l\ll n$ and is
widely used in theoretical analysis [4--8]. Integration of Eq. (2.2) for
motion between two subsequent passages at the aphelion (where $\dot x=0$ and
there is no energy exchange between the field and the atom) results to the
map (see [9--11] for details)
$$
\left\{
\begin{array}{ll}
\varepsilon _{j+1}= & \varepsilon _j-\pi F_0\varepsilon _0^2h\left(
\varepsilon _{j+1}\right) \sin \vartheta _j, \\
\vartheta _{j+1}= & \vartheta _j+2\pi \varepsilon _{j+1}^{-3/2}-\pi
F_0\varepsilon _0^2\eta \left( \varepsilon _{j+1}\right) \cos \vartheta _j
\end{array}
\right. \eqno{(2.3)}
$$
where
$$
h\left( \varepsilon _{j+1}\right) =\frac 4{\varepsilon _{j+1}}{\bf J}
_{s_{j+1}}^{\prime }(s_{j+1}).\eqno{(2.4)}
$$
Here $s\equiv \varepsilon ^{-3/2}=\omega /(-2E)^{3/2}=\omega /\Omega $ is
the relative frequency of the field, i.e., the ration of the field frequency
$\omega $ to the Kepler orbital frequency $\Omega =\left( -2E\right) ^{3/2}$
, and ${\bf J}_s^{\prime }(z)$ is the derivative of the Anger function with
respect to the argument $z$. The function $\eta \left( \varepsilon
_{j+1}\right) $ may be obtained from requirement of area-preserving of the
map (2.3)
$$
\frac{\partial \left( \varepsilon _{j+1},\vartheta _{j+1}\right) }{\partial
\left( \varepsilon _j,\vartheta _j\right) }=1.\eqno{(2.5)}
$$
This requirement yields
$$
\eta \left( \varepsilon _{j+1}\right) =\frac{dh\left( \varepsilon
_{j+1}\right) }{d\varepsilon _{j+1}}.\eqno{(2.6)}
$$
The derivative of the Anger function
$$
{\bf J}_s^{\prime }(s)=\frac 1\pi \int\limits_0^\pi \sin \left[ s\left(
x-\sin x\right) \right] \sin xdx\eqno{(2.7)}
$$
is a very simple analytical function which may be approximated quite well by
some combination [10] of expansion in powers of $s$
$$
{\bf J}_s^{\prime }(s)=\frac{1+\frac 5{24}s^2}{2\pi \left( 1-s^2\right) }
\sin \pi s,\quad s\leq 1\eqno{(2.8)}
$$
and of the asymptotic form
$$
{\bf J}_s^{\prime }(s)=\frac b{s^{2/3}}-\frac a{5s^{4/3}}-\frac{\sin \pi s}{
4\pi s^2},\quad s\gg 1\eqno{(2.9)}
$$
where
$$
a=\frac{2^{1/3}}{3^{2/3}\Gamma \left( 2/3\right) }\simeq 0.4473,\quad b=
\frac{2^{2/3}}{3^{1/3}\Gamma \left( 1/3\right) }\simeq 0.41085.\eqno{(2.10)}
$$
The map (2.3) is the general mapping form of the classical equations of
motion for the one-dimensional hydrogen atom in a microwave field derived in
the classical perturbation theory approximation. Some analytical and
numerical analysis of this map has been done in Refs. [9--11]. Here we
analyze different special cases of the map (2.3).

\subsection{\bf High frequency limit}

For the relatively high frequencies of the field, $s\gg 1$ ($s\geq 2$),
theoretical analysis of the classical dynamics of the one-dimensional
hydrogen atom in a microwave field is relatively simple. That is why, the
energy changes of the electron, $\left( E_{j+1}-E_j\right) $ and $\left(
\varepsilon _{j+1}-\varepsilon _j\right) $, do not depend on the initial
energy $\varepsilon _j$ and relative frequency $s\gg 1$. Indeed, using the
asymptotic form of the derivative of the Anger function, ${\bf J}_s^{\prime
}(s)=b/s^{2/3}$, we have $h\left( \varepsilon _{j+1}\right) =4b=const.$, $
\eta \left( \varepsilon _{j+1}\right) =0$ and, consequently, the following
map
$$
\left\{
\begin{array}{ll}
\varepsilon _{j+1}= & \varepsilon _j-4\pi bF_0\varepsilon _0^2\sin \vartheta
_j, \\
\vartheta _{j+1}= & \vartheta _j+2\pi \varepsilon _{j+1}^{-3/2}.
\end{array}
\right. \eqno{(2.11)}
$$
Note, that scaled classical dynamics according to maps (2.3) and (2.11)
depends only on single combination of the field parameters, i.e., on the
scaled field strength $F_s=F_0\varepsilon _0^2=F/\omega ^{4/3}$.

By the standard [12, 13] linearization procedure, $\varepsilon
_j=\varepsilon _0+\Delta \varepsilon _j$, in the vicinity of the integer
relative frequency (resonance), $s_0=\varepsilon _0^{-3/2}=m$ with $m$
integer, the map (2.11) may be transformed to the standard (Chirikov) map
$$
\left\{
\begin{array}{ll}
I_{j+1}= & I_j+K\sin \vartheta _j, \\
\vartheta _{j+1}= & \vartheta _j+I_{j+1}.
\end{array}
\right. \eqno{(2.12)}
$$
Here $I_j=-3\pi \Delta \varepsilon _j/\varepsilon _0^{5/2}$ and $K=12\pi
^2bF_0/\sqrt{\varepsilon _0}$.

From condition of the onset of classical chaos for the standard map, $K\geq
K_c\simeq 0.9816$ [12--15], we may, therefore, estimate the threshold field
strength for chaotization of dynamics and ionization of the atom in the high
frequency field
$$
F_0^c=K_c/\left( 12\pi ^2bs_0^{1/3}\right) \simeq 0.02s_0^{-1/3}.
\eqno{(2.13)}
$$
Some times [6] one writes the map (2.11) for a variable $N=-1/2n^2\omega $,
the change of which gives the number of absorbed photons,
$$
\left\{
\begin{array}{ll}
N_{j+1}= & N_j+2\pi \left( F/\omega ^{5/3}\right) \sin \vartheta _j, \\
\vartheta _{j+1}= & \vartheta _j+2\pi \omega \left( -2\omega N_{j+1}\right)
^{-3/2}.
\end{array}
\right. \eqno{(2.14)}
$$
We see that for such variables dynamics of the system depends on two
parameters: on the quantum scaled field strength $F_q=F/\omega ^{5/3}$ [2,
3] and on the field frequency $\omega $. Map (2.14) is, therefore, not the
most convenient one for analysis of the {\it classical} dynamics.

In general there are, however, no essential difficulties in theoretical
analysis of classical nonlinear dynamics of the highly excited hydrogen atom
in the microwave field of relative frequency $s_0=\omega n_0^3\geq 0.5$ when
the field strength is lower or comparable with the threshold field strength
for the onset of classical chaos, i.e., if the microwave field is
considerably weaker than the characteristic Coulomb field. In such a case,
energy change of the electron during the period of intrinsic motion is
relatively small and application of the classical perturbation theory for
derivation of the Kepler map (2.3) is sufficiently correct. Further analysis
of transition to chaotic behavior and of the ionization process may be based
on the map (2.3) and for $s_0\simeq 0.3\div 1.5$ results in the impressive
agreement between measured ionization curves and those obtained from the map
(2.3). Even analytical estimation of the threshold field strengths based on
this map is rather proper [9--11].

Sufficiently more complicated is analysis of transition to stochastic motion
and of ionization process in the region of low relative frequencies, $
s_0\leq 0.3$.

\subsection{\bf Low frequency limit\ }

For the low relative frequencies of the microwave field, $s\ll 1$, the map
(2.3) may be simplified as well. Using expansion of the function ${\bf J}
_s^{\prime }(s)$ in powers of $s$, ${\bf J}_s^{\prime }(s)\simeq s/2$, for $
s\ll 1$ we have according to Eqs. (2.4) and (2.6)
$$
\left\{
\begin{array}{ll}
h\left( \varepsilon _{j+1}\right) & =2/\varepsilon _{j+1}^{5/2} \\
\eta \left( \varepsilon _{j+1}\right) & =-5/\varepsilon _{j+1}^{7/2}.
\end{array}
\right. \eqno{(2.15)}
$$
Consequently map (2.3) transforms to the form
$$
\left\{
\begin{array}{ll}
\varepsilon _{j+1}= & \varepsilon _j-2\pi F_0\left( \varepsilon
_0^2/\varepsilon _{j+1}^{5/2}\right) \sin \vartheta _j, \\
\vartheta _{j+1}= & \vartheta _j+2\pi /\varepsilon _{j+1}^{3/2}+5\pi
F_0\left( \varepsilon _0^2/\varepsilon _{j+1}^{7/2}\right) \cos \vartheta
_j.
\end{array}
\right. \eqno{(2.16)}
$$
This map is a little bit more complicated than map (2.11) for high
frequencies, however, it may easily be analyzed as numerically as well as
analytically. Note first of all, that energy change of the electron during
the period of intrinsic motion (after one step of iteration), $\left|
\varepsilon _{j+1}-\varepsilon _j\right| ,$ is considerably smaller than the
binding energy of the electron $\varepsilon _j\simeq \varepsilon _0$ if the
field strength is lower or comparable with the threshold field strength,
i.e., $2\pi F_0\left( \varepsilon _0^2/\varepsilon _{j+1}^{5/2}\right)
\simeq 2\pi F_0\varepsilon _0^{-1/2}\ll \varepsilon _0$, or $2\pi F_0s_0\ll
1 $ if $F_0\leq F_0^{st}\simeq 0.13$ and $s_0\ll 1$. This indicates that the
map (2.16) is probably suitable for description of dynamics even in the low
frequency region where the field is relatively strong.

\subsubsection{\bf Adiabatic ionization}

For low frequencies, $2\pi s=2\pi /\varepsilon ^{3/2}\ll 1$, according to
the second equation of map (2.16) the change of the angle $\vartheta $ after
one step of iteration is small. As it was noticed above, the energy change
is relatively small too. Therefore, we may transform difference equations
(2.16) to differential equations of the form
$$
\left\{
\begin{array}{ll}
\frac{d\varepsilon }{dj}= & -
\frac{2\pi \varepsilon _0^2F_0}{\varepsilon ^{5/2}}\sin \vartheta , \\ \frac{
d\vartheta }{dj}= & \frac{2\pi }{\varepsilon ^{3/2}}+\frac{5\pi \varepsilon
_0^2F_0}{\varepsilon ^{7/2}}\cos \vartheta .
\end{array}
\right. \eqno{(2.17)}
$$
Dividing second equation of the system (2.17) by the first one we obtain one
differential equation
$$
\frac{d\left( \cos \vartheta \right) }{d\varepsilon }=\frac \varepsilon
{\varepsilon _0^2F_0}+\frac{5\cos \vartheta }{2\varepsilon }.\eqno{(2.18)}
$$
Analytical solution of Eq. (2.18) with the initial condition $\varepsilon
=\varepsilon _0$ when $\vartheta =\vartheta _0$ is
$$
\cos \vartheta =z^5\cos \vartheta _0-2z^4\left( 1-z\right) /F_0,\quad z=
\sqrt{\varepsilon /\varepsilon _0}.\eqno{(2.19)}
$$
Eq. (2.19) describes motion of the system in $\varepsilon $ and $\vartheta $
variables, i.e., represents functional dependence between two dynamical
variables. For relatively low values of $F_0$, i.e., for $F_0<\frac
25z^4=\frac 25\left( \frac \varepsilon {\varepsilon _0}\right) ^2$, the
right-hand side of Eq. (2.18) is positive for all phases $\vartheta $.
Therefore, $\cos \vartheta $ and $\varepsilon $ decrease and increase
simultaneously and, according to Eq. (2.19), there is a motion in all
interval $\left[ 0,2\pi \right] $ of the angle $\vartheta .$ For $F_0>\frac
25z^4$, however, the increase of the angle $\vartheta $ in the interval $
0\div \pi $ turns at $\vartheta \simeq \pi $ into the decrease. This results
in fast decrease of $\varepsilon $ and to ionization process. It is easy to
understand from analysis of Eq. (2.19) that the minimal value of $F_0$ for
such a motion (resulting in ionization) corresponds to $\vartheta _0=0$ and $
\vartheta =\pi $. This value of $F_0$ is very close to the maximal value of $
F_0$ resulting to the motion in all interval $\left[ 0,2\pi \right] $ of $
\vartheta $, i.e., the maximum of the expression
$$
F_0=2z\left( 1-z\right) /\left( 1+z^5\right) .\eqno{(2.20)}
$$
This maximum is at $z=z_0$, where $z_0$ is a solution of the equation $
z^5+5z-4=0$, being $z_0\simeq 0.75193$. The critical value of the relative
field strength, therefore, is $F_0^0=2z_0^4/5=0.1279$ which is only 1\
lower the adiabatic ionization threshold $F_0^{st}=2^{10}/\left( 3\pi
\right) ^4=0.1298$.

\subsubsection{\bf Chaotic ionization}

For higher relative frequencies, $s_0\geq 0.1$, ionization process is due to
chaotic dynamics of the highly excited electron of the hydrogenic atom in a
microwave field. There are different criterions for estimation of the
parameters when dynamics of the nonlinear system becomes chaotic. For
analysis of transition to chaotic behavior of the motion described by maps
(2.3), (2.11) and (2.16) the most proper, to is the criterion related with
the chaotization of the phases [13]
$$
K=\max \left| \frac{\delta \vartheta _{j+1}}{\delta \vartheta _j}-1\right|
\geq 1.\eqno{(2.21)}
$$
Here $\max $ means the maximum with respect to the phase $\vartheta _j$ and
variation of the phase $\vartheta _{j+1}$ with respect to the phase $
\vartheta _j$ means the full variation including dependence of $\vartheta
_{j+1}$ on $\vartheta _j$ through the variable $\varepsilon _{j+1}$ in Eqs.
(2.3), (2.11) and (2.16).

Applying criterion (2.21) to the general map (2.3)--(2.4) we obtain the
threshold field strength
$$
F_0^c=\frac{\varepsilon ^{7/2}}{12\pi ^2\varepsilon _0^2{\bf J}_s^{\prime
}(s)}.\eqno{(2.22)}
$$
If $\varepsilon \simeq \varepsilon _0$ Eq. (2.22) yields the result
$$
F_0^c=\left( 12\pi ^2s{\bf J}_s^{\prime }(s)\right) ^{-1}\eqno{(2.23)}
$$
which for $s\gg 1$ coincides with Eq. (2.13).

For more precise evaluation of the critical field strengths we should take
into account the change (increase) of the electron's energy due to the
influence of the electromagnetic field [11]. For higher relative frequency $
s $ or lower scaled energy $\varepsilon _j$ the threshold ionization field
is lower. Therefore, if the scaled energy $\varepsilon _j$ decreases in a
result of relatively regular dynamics in not very strong microwave field,
then it is sufficient the lower field strength for transition to the chaotic
dynamics.

\section{Transition to nonchaotic dynamics and synchronization in randomly
driven systems}

When an ensemble of bounded in a fixed external potential particles with
different initial conditions are driven by an identical sequence of random
forces, the ensemble of trajectories may become identical at long times,
i.e. synchronization of the identical systems by common noise may be
observed. Fahy and Hamann [16] considered a particle of mass $m$ moving
according to Newton's equations in a potential $V(x)$, except that at
regular time intervals $\tau $ the particle is stopped and its velocity is
reset to random value chosen from a Maxwell distribution with temperature $T$
. It should be stressed that for every particle of the ensemble it was given
an identical, randomly chosen velocity at the start of each step of time
length $\tau $. This motion is in many respects similar to Brownian motion
of the particles at a temperature $T$. However, if the time interval $\tau $
between stops is lower than a threshold value $\tau _c$, the final
trajectories of the particles are independent on the initial conditions; all
trajectories become point by point identical in time. Although the
trajectory is highly erratic and random, the system is not chaotic.

The similar effect may also be observed in a more general and realistic
(from the physical point of view) case, i.e., when mixing at time intervals $
\tau $ some part $\alpha $ of the old velocity ${\bf v}^{old}$ with random
velocity ${\bf v}^{ran}$ to get a new starting velocity ${\bf v}
^{new}=\alpha {\bf v}^{old}+{\bf v}^{ran}$. Here a threshold value $\tau _c$
depends on $\alpha $.

Let us consider a particle of mass $m$ moving in a one-dimensional potential
$V(x)$ which confines particles to a finite region. At a time intervals $
\tau _i$ the particle is partially stopped and its velocity is reset to a
new starting velocity $v_i=\alpha v_i^{old}+v_i^{ran}$. Between the stops
the particle moves according to Newton's equations
$$
{\frac{dx}{{dt}}}=v,~~~~~~{\frac{d^2x}{{dt^2}}}=-{\frac 1{{m}}}{\frac{dV}{{dx
}}}.\eqno(3.1)
$$
When two particles initially at points $x_0$ and $x_0^{\prime }$ are started
with velocities $v_0$ and $v_0^{\prime }$ and are driven by an identical
sequence of random velocities $v_i^{ran}$ at the same time intervals $\tau
_i $, coordinates and velocities of them may accidentally draw closer to one
another. The convergence of the two trajectories to the single final
trajectory will depend on the evolution with a time of the small variances
of the distance $\Delta x_i=x_i^{\prime }-x_i$ and velocity $\Delta
v_i=v_i^{\prime }-v_i$. Moreover, we investigate a transition from chaotic
to nonchaotic behavior. Generally, such a transition may be detected from
analysis of behavior of the neighboring trajectories and it is described by
the Lyapunov characteristic exponents and $KS$ metric entropy of the flow of
trajectories in a given region of phase space [12, 13, 17--20].

From formal solutions $x=x(x_i,v_i,t)$ and $v=v(x_i,v_i,t)$ of equations
(3.1) with initial conditions $x=x_i$ and $v=v_i$ at $t=0$ it follows an
equation for $\Delta x(t)$ and $\Delta v(t)$ at a time moment $t$:
$$
\pmatrix{\Delta x(t)\cr \Delta v(t)}={\bf T}(\alpha ;x_i,v_i,t)
\pmatrix{\Delta x_i\cr \Delta v_i}\eqno(3.2)
$$
where the matrix ${\bf T}$ is of the form
$$
{\bf T}=\pmatrix{T_{xx}&\alpha T_{xv}\cr
T_{vx}&\alpha T_{vv}\cr}=
\pmatrix{\displaystyle{\partial x(x_i,v_i,t)\over\partial x_i}&
\displaystyle{\alpha{\partial x(x_i,v_i,t)\over\partial v_i}}\cr
\displaystyle{\partial v(x_i,v_i,t)\over\partial x_i}&
\displaystyle{\alpha{\partial v(x_i,v_i,t)\over\partial v_i}}\cr}.\eqno(3.3)
$$
Note, that the similar method of investigation is used in the theory of
transition to chaos in classical systems [17, 18]. However, the motion in
the form (3.2) and (3.3) is represented as the non-area-preserving tangent
map, while classical dynamics of the conservative systems may be represented
by the area-preserving maps.

According to equations (3.1) and (3.3) matrix elements $T_{xx}$ and $T_{xv}$
satisfy the equation
$$
{\frac{d^2T_x}{dt^2}}=-{\frac 1m}{\frac{d^2V}{dx^2}}\Biggl|
_{x=x(x_i,v_i,t)}T_x\eqno(3.4)
$$
while $T_{vx}=\dot T_{xx},~T_{vv}=\dot T_{xv}$ and the initial conditions at
$t=0$ are:
$$
T_{xx}(x_i,v_i,0)=T_{vv}=1,~~~T_{xv}=T_{vx}=0,
$$
$$
\dot T_{xx}(x_i,v_i,0)=\dot T_{vv}=0,~\dot T_{xv}=1,~\dot T_{vx}=-{\frac 1m}{
\frac{d^2V}{dx^2}}\Biggl|_{x=x_i}.\eqno(3.5)
$$
Therefore, the dynamics of the distance between the particles $\Delta x$ and
the difference of the velocity $\Delta v$ may be represented by the
non-area-preserving mapping form of the equations of motion
$$
\pmatrix{\Delta x_{i+1}\cr \Delta v_{i+1}}={\bf T}(\alpha ;x_i,v_i,\tau _i)
\pmatrix{\Delta x_i\cr \Delta v_i}.\eqno(3.6)
$$
In general, the intervals between the resets of the velocity $\tau _i$ may
be depending on the number of step $i$.

Further analysis of the model may be based on the general theory of the
dynamics of classical systems represented as maps [17--19]. Thus, for $
\alpha =0$ the Lyapunov exponent is defined as
$$
\lambda ={\lim \limits_{N\to \infty }}{\frac 1N}\sum\limits_{i=1}^N{\frac
1{\tau _i}}\ln \Bigl|{T_{xx}(x_i,v_i,\tau _i)}\Bigr|\eqno(3.7)
$$
and may be easily evaluated numerically.

For $\alpha =1$ the map (3.6) is area-preserving and $\det {\bf T}
(1;x_i,v_i,\tau _i)=1$, while in general $\det {\bf T}=\alpha $, $Tr{\bf T}
=T_{xx}+\alpha T_{vv}$ and the eigenvalues $\mu _{1,2}$ of the ${\bf T}$
matrix are given by equation
$$
\mu ^2-\mu Tr{\bf T}+\det {\bf T}=0
$$
which yields
$$
\mu _{1,2}={\frac 12}\Bigl[T_{xx}+\alpha T_{vv}\mp \sqrt{(T_{xx}+\alpha
T_{vv})^2-4\alpha }\Bigr].
$$
So, the eigenvalues come in reciprocal pair, $\mu _1\mu _2=\alpha $. For $
(T_{xx}+\alpha T_{vv})^2-4\alpha <0$ the eigenvalues form a complex
conjugate pair with $|\mu _1|=|\mu _2|=\sqrt{\alpha }$, otherwise the
eigenvalues are real.

Generally, the mapping ${\bf T}(\alpha ;x_i,v_i,\tau _i)$ in (3.6) is
depending on the starting coordinates $x_i$ and $v_i$. Therefore,
calculation of the mapping for $n$ steps, ${\bf T}_n={\bf T}(\alpha
;x_{i+n-1},v_{i+n-1},\tau _{i+n-1})\cdot {\bf T}(\alpha
;x_{i+n-2},v_{i+n-2},\tau _{i+n-2})~\cdot \cdot \cdot ~{\bf T}(\alpha
;x_i,v_i,\tau _i)$, and of the corresponding eigenvalues are complicated
problems. Further we will evaluate the averaged quantities
$$
\sigma _{1,2}=\bigl\langle{\frac 1{\tau _i}}\ln |\mu _{1,2}|\bigr\rangle
=\lim \limits_{N\to \infty }{\frac 1N}\sum\limits_{i=1}^N{\frac 1{\tau _i}}
\ln \Bigl|\mu _{1,2}(x_i,v_i,\tau _i)\Bigr|\eqno(3.8)
$$
which are analogous of the averaged Lyapunov exponent (3.7), characterize
the rate of the exponential increase of the separation of the two initially
adjacent points and are related with the $KS$ entropy of the system [17,
18]. Comparisons of the threshold values $\tau _c$ from the direct numerical
simulations with those from the criterion
$$
\sigma _{largest}=0\eqno(3.9)
$$
indicate to the usefulness of the quantities (3.8) for analysis of
transition from nonchaotic to chaotic behavior and synchronization of the
systems [21--23].

Therefore, theoretical analysis based on the mapping form of equations of
motion for the distance between the particles and the difference of the
velocity allows to simplify the problem of investigation of transition to
nonchaotic behavior and results in the expressions for the criteria of the
nonchaotic motion. Theoretical results agree well with the direct numerical
simulations and indicate to the possibilities of generalization of the
model, e.g. to more degrees of freedom, for random values of the time
intervals between the resets of the velocity and for systems driven by the
random forces [21--23]. In paper [24] this method has been generalized and
used for analysis of systems with repulsive force between particles, some
scaling properties for the threshold reset time have been derived and it has
been suggested that such convergence of chaotic orbits is a rather general
phenomenon.

\section{Dynamics of quantum systems}

\subsection{\bf Two-level system}

Let's consider the simplest quantum dynamical process and the influence of
frequent measurements on the outcome of the dynamics. Time evolution of the
amplitudes $a_1(t)$ and $a_2\left( t\right) $ of the two-state wave function
$$
\Psi =a_1(t)\Psi _1+a_2\left( t\right) \Psi _2\eqno{(4.1)}
$$
of the system in the resonance field (in the rotating wave approximation) or
of the spin-half system in a constant magnetic field may be represented as
$$
\left\{
\begin{array}{ll}
a_1(t)= & a_1(0)\cos \frac 12\Omega t+ia_2(0)\sin \frac 12\Omega t \\
a_2(t)= & ia_1(0)\sin \frac 12\Omega t+a_2(0)\cos \frac 12\Omega t,
\end{array}
\right. \eqno{(4.2)}
$$
where $\Omega $ is the Rabi frequency. We introduce a matrix ${\bf A}$
representing time evolution during the time interval $\tau $ (between time
moments $t=k\tau $ and $t=(k+1)\tau $ with integer $k$) and rewrite Eq.
(4.2) in the mapping form
$$
\left( \matrix{a_1(k+1)\cr a_2(k+1)}\right) ={\bf A}\left(
\matrix{a_1(k)\cr a_2(k)}\right) \eqno(4.3)
$$
where the evolution matrix ${\bf A}$ is given by
$$
{\bf A}=\left(
\matrix{\cos\varphi& i\sin\varphi\cr
i\sin\varphi & \cos\varphi}\right) ,~~~\varphi ={\frac 12}\Omega \tau .
\eqno(4.4)
$$
Evidently, the evolution of the amplitudes from time $t=0$ to $t=T=n\tau $
may be expressed as
$$
\left( \matrix{a_1(n)\cr a_2(n)}\right) ={\bf A}^n\left(
\matrix{a_1(0)\cr
a_2(0)}\right) .\eqno(4.5)
$$
One may calculate matrix ${\bf A}^n$ by the method of diagonalization of the
matrix ${\bf A.}$ The result naturally is
$$
{\bf A}^n=\left(
\matrix{\cos n\varphi & i\sin n\varphi\cr
i\sin n\varphi & \cos n\varphi}\right) .\eqno(4.6)
$$
Note that $n\varphi =\frac 12\Omega T.$

Equations (4.2)--(4.6) represent time evolution of the system without the
intermediate measurements in the time interval $0\div T$. If at $t=0$ the
system was in the state $\Psi _1$, i.e. $a_1(0)=1$ and $a_2(0)=0$, and if $
\Omega T=\pi $ then at the time moment $t=T$ we would certainly find the
system in the state $\Psi _2$, i.e. it would be $\left| a_1(T)\right| ^2=0$
and $\left| a_2(T)\right| ^2=1$, a certain transition between the states.

Let's consider now the dynamics of the system with the intermediate
measurements every time interval $\tau $. Measurement of the system's state
in the time moment $t=k\tau $ projects the system into the state $\Psi _1$
with the probability $p_1(k)=\mid a_1(k)\mid ^2$ or into the state $\Psi _2$
with the probability $p_2(k)=\mid a_2(k)\mid ^2$. After the measurement we
know the probabilities $p_1(k)$ and $p_2(k)$ but we have no information
about the phases $\alpha _1(k)$ and $\alpha _2(k)$ of the amplitudes
$$
a_1(k)=\left| a_1(k)\right| e^{i\alpha _1(k)},~~~a_2(k)=\left| a_2(k)\right|
e^{i\alpha _2(k)},\eqno{(4.7)}
$$
i.e. the phases $\alpha _1(k)$ and $\alpha _2(k)$ after every act of the
measurement are random [25]. This results in the equation for the
probabilities
$$
\left( \matrix{p_1(k+1)\cr p_2(k+1)}\right) ={\bf M}\left(
\matrix{p_1(k)\cr
p_2(k)}\right) ,\eqno{(4.8)}
$$
where
$$
{\bf M}=\left(
\matrix{\cos^2\varphi & \sin^2\varphi\cr
\sin ^2\varphi & \cos ^2\varphi}\right) \eqno(4.9)
$$
is the evolution matrix for the probabilities. The full evolution from the
initial time $t=0$ until $t=T$ with the $(n-1)$ equidistant intermediate
measurement is described by the equation
$$
\left( \matrix{p_1(n)\cr p_2(n)}\right) ={\bf M}^n\left(
\matrix{p_1(0)\cr
p_2(0)}\right) .\eqno(4.10)
$$
The result of calculation of the matrix ${\bf M}^n$ by the method of
diagonalization of the matrix ${\bf M}$ is
$$
{\bf M}^n={\frac 12}\left(
\matrix{1+\cos ^n2\varphi & 1-\cos ^n2\varphi\cr
1-\cos ^n2\varphi & 1+\cos ^n2\varphi}\right) .\eqno(4.11)
$$
From Eqs. (4.10) and (4.11) we recover the quantum Zeno effect [25-27]: if
initially the system is in the state $\Psi _1$, than the result of the
evolution until the time moment $T=n\tau =\pi /\Omega $ (after the $\pi $
-pulse) with the $(n-1)$ intermediate measurement will be characterized by
the probabilities $p_1(T)$ and $p_2(T)$ for finding the system in the states
$\Psi _1$ and $\Psi _2$ respectively:
$$
\left\{
\begin{array}{ll}
p_1(T)= & \frac 12(1+\cos ^n2\varphi )\simeq \frac 12(1+e^{-
\frac{\pi ^2}{2n}})\simeq 1-\frac{\pi ^2}{4n}\rightarrow 1, \\ p_2(T)= &
\frac 12(1-\cos ^n2\varphi )\simeq \frac 12(1-e^{-\frac{\pi ^2}{2n}})\simeq
\frac{\pi ^2}{4n}\rightarrow 0,~n\rightarrow \infty .
\end{array}
\right. \eqno(4.12)
$$
We see that results of equations (4.10)--(4.12) represent the inhibition of
the quantum dynamics by measurements and coincide with those obtained by the
density matrix technique [26, 27]. This also confirms correctness of the
proposition that the act of the measurement may be represented as
randomization of the amplitudes' phases. Further we will use this
proposition and the same method for the analysis of the repeated measurement
influence for the quantum dynamics of multilevel systems which classical
counterparts exhibit chaos. We restrict ourselves to the strongly driven by
a periodic force systems with one degree of freedom. The investigation is
also based on the mapping equations of motion for such systems.

\subsubsection{\bf Dynamics of multilevel systems}

In general the classical equations of motion are nonintegrable and the
Schr\"odinger equation for strongly driven systems may not be solved
analytically. However, mapping forms of the classical and quantum equations
of motion greatly facilitates the investigation of stochasticity and
quantum--classical correspondence for the chaotic dynamics. From the
standpoint of an understanding of the manifestation of the measurements for
the dynamics of the multilevel systems the region of large quantum numbers
is of greatest interest. Here we may use the quasiclassical approximation
and convenient variables are the angle $\theta $ and the action $I$. One of
the simplest systems in which the dynamical chaos and its quantum
localization may be observed is a system with one degree of freedom
described by the unperturbed Hamiltonian $H_0(I)$ and driven by periodic
kicks. The full Hamiltonian $H$ of the driven system may be represented as
$$
H(I,\theta ,t)=H_0(I)+k\cos \theta \sum_j\delta \left( t-j\tau \right)
\eqno{(4.13)}
$$
where $\tau $ and $k$ are the period and strength of the perturbation,
respectively.

For the derivation of the quantum equations of motion we expand the state
function $\psi (\theta ,t)$ of the system through the quasiclassical
eigenfunctions, $\varphi _n(\theta )=e^{in\theta }/\sqrt{(}2\pi )$, of the
Hamiltonian $H_0$,
$$
\psi (\theta ,t)=(2\pi )^{-1/2}\sum\limits_na_n(t)i^{-n}e^{-in\theta }.
\eqno(4.14)
$$
Here the phase factor $i^{-n}$ is introduced for the maximal simplification
of the quantum map. Integrating the Schr\"odinger equation over the period $
\tau $, we obtain the following maps for the amplitudes before the
appropriate kicks
$$
a_m(t_{j+1})=e^{-iH_0(m)\tau }\sum\limits_na_n(t_j)J_{m-n}(k),~~~t_j=j\tau
\eqno(4.15)
$$
where $J_m(k)$ is the Bessel function.

The form (4.15) of the map for the quantum dynamics is rather common:
similar maps may be derived for the monochromatic perturbations as well,
e.g. for an atom in a microwave field [28]. A particular case of map (4.15),
corresponding to the model of a quantum rotator $H=I^2/2$, has been
comprehensively investigated with the aim of determining the relationship
between classical and quantum chaos. It has been shown that under the onset
of dynamical chaos at $K\equiv \tau k>K_c=0.9816$, motion with respect to $I$
is not bounded and it is of a diffusion nature in the classical case, while
in the quantum description diffusion with respect to $m$ is bounded, i.e.
the diffusion ceases after some time and the state of the system localizes
exponentially. This phenomenon turns out to be typical for models (4.15)
with nonlinear Hamiltonians $H_0(I)$ and for other quantum systems. The
quantum interference effect is essential for such dynamics and it results in
the quantum evolution being quantitatively different from the classical
motion. Quantum equations of motion, i.e. the Schr\"odinger equation and the
maps for the amplitudes, are linear equations with respect to the wave
function and probability amplitudes. Therefore, the quantum interference
effect manifests itself even for quantum dynamics of the systems, the
classical counterparts of which are described by nonlinear equations;
chaotic dynamics of the later exhibit a dynamical chaos. On the other hand,
quantum equations of motion are very complex as well. Thus, the
Schr\"odinger equation is a partial differential equation with the
coordinate and time dependent coefficients, while the system of equations
for the amplitudes is the infinite system of equations. Moreover, for the
nonlinear Hamiltonian $H_0(m)$ the phases' increments, $H_0(m)\tau $, during
the free motion between two kicks while reduced to the basic interval $
\left[ 0,2\pi \right] $ are the pseudorandom quantities as a function of the
state's quantum number $m$. This causes a very complicated and irregular
quantum dynamics of the classically chaotic systems. We observe not only
very large and apparently irregular fluctuations of probabilities of the
states' occupation but also almost irregular fluctuations in time of the
momentum dispersion [25, 28].

However, the quantum dynamics of such driven by the external periodic force
systems is coherent and the evolution of the amplitudes $a_m(t_{j+1})$ in
time is linear: they are defined by the linear map (4.15) with the time
independent coefficients. The nonlinearity of the Hamiltonian $H_0(I)$,
being the reason of the classical chaos, causes the pseudorandom nature of
the increments of the phases, $H_0(m)\tau $, as a function of the state's
number $m$ (but constant in time). These increments of the phases remain the
same for each kick. So, the dynamics of the amplitudes $a_m(t_{j+1})=\left|
a_m(t_{j+1})\right| e^{i\alpha _m\left( t_{j+1}\right) }$ and of their
phases, $\alpha _m\left( t_{j+1}\right) $, is strongly deterministic and
non-chaotic but very complicated and apparently irregular. For instance, the
phases $\alpha _m\left( t_{j+1}\right) $ are phases of the complex
amplitudes, $a_m(t_{j+1})$, which are linear combinations (4.15) of the
complex amplitudes, $a_n(t_j)$, before the preceding kick with the
pseudorandom coefficients, $e^{-iH_0(m)\tau }J_{m-n}(k)$. Nevertheless, the
iterative equation (4.15) is a {\it linear transformation with coefficients
regular in time.} That is why, we observe for such dynamics the
quasiperiodic reversible in the time evolution [28] with the quantum
localization of the pseudochaotic motion.

In paper [28] it has been demonstrated that this peculiarity of the
pseudochaotic quantum dynamics is indeed due to the pseudorandom nature of
the phases, $H_0(m)\tau $, in Eq. (4.15) as a function of the eigenstate's
quantum number $m$ (but not of the kick's number $j$). Replacing the
multipliers $\exp \left[ -iH_0(m)\tau \right] $ in Eq. (4.15) by the
expressions $\exp \left[ -i2\pi g_m\right] $, where $g_m$ is a sequence of
random numbers that are uniformly distributed in the interval $\left[
0,1\right] $, we observe the quantum localization as well [28]. The
essential point here is the independence of the phases $H_0(m)\tau $ or $
2\pi g_m$ on the step of iteration $j$ or time $t$.

\subsubsection{\bf Influence of repetitive measurement on the quantum
dynamics}

Each measurement of the system's state projects the state into one of the
energy state $\varphi _m$ with the definite $m$. Therefore, if we make a
measurement of the system after the kick $j$ but before the next $(j+1)$
kick we will find the system in the states $\varphi _m$ with the appropriate
probabilities $p_m(j)=\left| a_m(t_j)\right| ^2$.

In the calculations of the system's dynamics the influence of the
measurements may be taken into account through randomization of phases of
the amplitudes after the measurement of the appropriate state's populations.
The phases of amplitudes after the measurement are completely random and
uncorrelated with the phases before the measurement, after another
measurements and with the phases of other measured or unmeasured states.
Therefore, after the full measurement of the system after the kick $j$, all
phases of the amplitudes $a_m(t_j)$ are random. So, this full measurement of
the system's state influences on the further dynamics of the system through
the randomization of the phases of amplitudes. This fact may be expressed by
replacement in Eqs. (4.15) of the amplitudes $a_m(t_{j+1})$ by the
amplitudes $e^{i\beta _m(t_{j+1})}a_m(t_{j+1})$ with the random phases $
\beta _m(t_{j+1})$. The essential point here is that the phases $\beta
_m(t_{j+1})$ are different, uncorrelated for the different measurements,
i.e. for different time moments of the measurement $t_{j+1}$. This is the
principal difference of the random phases $\beta _m(t_{j+1})$ from the
phases $H_0(m)\tau $ in Eqs. (4.15) which are pseudorandom variables as
functions of the eigenstate's quantum number $m$ (but not of the time moment
$t_{j+1}$).

Instead of representing the detailed quantum dynamics expressed as the
evolution of all amplitudes in the expansion of the wave function (4.14) we
may represent only dynamics of the momentum dispersion $\left\langle
(m_j-m_0)^2\right\rangle =\sum\limits_m\left( m-m_0\right) ^2\left|
a_m\left( t_j\right) \right| ^2$ where $m_0$ is the initial momentum
(quantum number). Such a representation of the dynamics is simpler, more
picturesque and more comfortable for comparison with the classical dynamics.

Theoretically differences of dynamics without measurement and with the
measurement may be understood from the iterative equations for the momentum
dispersion:
$$
\left\langle (m_{j+1}-m_0)^2\right\rangle =\sum\limits_m\left( m-m_0\right)
^2\left| a_m\left( t_{j+1}\right) \right| ^2,\eqno{(4.16)}
$$
where
$$
\left| a_m\left( t_{j+1}\right) \right| ^2=\sum_{n,n^{^{\prime
}}}J_{m-n}\left( k\right) J_{m-n^{^{\prime }}}\left( k\right) a_n\left(
t_j\right) a_{n^{^{\prime }}}^{*}\left( t_j\right) .\eqno{(4.17)}
$$
Substitution of Eq. (4.17) into Eq. (4.16) yields
$$
\begin{array}{c}
\left\langle (m_{j+1}-m_0)^2\right\rangle =\sum\limits_{m,n}\left(
m-m_0\right) ^2J_{m-n}^2\left( k\right) \left| a_n\left( t_j\right) \right|
^2 \\
+2\sum\limits_{m,n}\sum\limits_{n^{^{\prime }}<n}\left( m-m_0\right)
^2J_{m-n}\left( k\right) J_{m-n^{^{\prime }}}\left( k\right) Re\left[
a_n\left( t_j\right) a_{n^{^{\prime }}}^{*}\left( t_j\right) \right] .
\end{array}
\eqno{(4.18)}
$$

For the random phase differences of the amplitudes $a_n\left( t_j\right) $
and $a_{n^{^{\prime }}}^{*}\left( t_j\right) $ with $n^{^{\prime }}\neq n$,
which is a case after the measurement of the system's state, the second term
of Eq. (4.18) on the average equals zero. Then from Eq. (4.18) we have
$$
\begin{array}{c}
\left\langle (m_{j+1}-m_0)^2\right\rangle =\sum\limits_n\left| a_n\left(
t_j\right) \right| ^2\sum\limits_m\left( m-m_0\right) ^2J_{m-n}^2\left(
k\right) \\
=\sum\limits_m\left| a_m\left( t_j\right) \right| ^2\left( m^2-m_0^2+\frac{
k^2}2\right) =\left\langle (m_j-m_0)^2\right\rangle +\frac{k^2}2.
\end{array}
\eqno{(4.19)}
$$
In the derivation of Eq. (4.19) we have used the summation expressions
$$
\sum\limits_mmJ_{m-n}^2\left( k\right) =0\quad \mbox{and}\quad
\sum\limits_mm^2J_{m-n}^2\left( k\right) =n^2+\frac{k^2}2.\eqno{(4.20)}
$$

Therefore, according to Eq. (4.19) for the uncorrelated phases of the
amplitudes $a_n\left( t_j\right) $ and $a_{n^{^{\prime }}}^{*}\left(
t_j\right) $ with $n^{^{\prime }}\neq n$ the dispersion of the momentum as a
result of every kick increases on the average in the magnitude $k^2/2$, the
same as for the classical dynamics. Thus, we reveal that repetitive
measurement of the multilevel systems with quantum suppression of classical
chaos results in the delocalization of the states superposition and
restoration of the chaotic dynamics. Since this effect is reverse to the
quantum Zeno effect we have called this phenomenon the {\it 'quantum
anti-Zeno effect'} [25].

\section{Conclusions}

In this paper on the concrete examples we have demonstrated usefulness of
the mapping equations for analysis of dynamics the nonlinear systems:
transition to chaotic behavior in Hamiltonian systems, synchronization of
chaotic systems driven by identical noise and effect of repetitive
measurements for quantum dynamics.

\vspace{0.5cm}

{\Large {\bf References }}

\begin{enumerate}
\item  L. D. Landau and E. M. Lifshitz, {\it Classical Field Theory}
(Pergamon, New York, 1975).

\item  B. Kaulakys, V. Gontis, G. Hermann, and A. Scharmann, Phys. Lett. A
{\bf 159}, 261 (1991); B. Kaulakys, Acta Phys. Pol. B {\bf 23}, 313 (1992).

\item  B. Kaulakys, V. Gontis, and G. Vilutis, Lith. Phys. J. (Allerton
Press, Inc.) {\bf 33}, 290 (1993); B. Kaulakys and G. Vilutis, in {\it AIP
Conf. Proc.} (AIP, New York) {\bf 329}, 389 (1995); e-print archives:
xxx.lanl.gov/quant-ph/9504007.

\item  N. B. Delone, V. P. Krainov, and D. L. Shepelyansky, Usp. Fiz. Nauk
{\bf 140}, 355 (1983) [Sov. Phys.-Usp. {\bf 26}, 551 (1983)].

\item  G. Casati, B. V. Chirikov, D. L. Shepelyansky, and I. Guarneri, Phys.
Rep. {\bf 154}, 77 (1987).

\item  G. Casati, I. Guarneri, and D. L. Shepelyansky, IEEE J. Quantum
Electron. {\bf 24}, 1420 (1988).

\item  R. V. Jensen, S. M. Susskin, and M. M. Sanders, Phys. Rep. {\bf 201},
1 (1991).

\item  P. M. Koch, in {\it Chaos and Quantum Chaos}, edited by W. Heiss,
Lecture Notes in Physics Vol. {\bf 411} (Springer-Verlag, Berlin, 1992), p.
167; P. M. Koch and K. A. H. van Leeuwen, Phys. Rep. {\bf 255}, 289 (1995).

\item  V. Gontis and B. Kaulakys, Deposited in VINITI as No.5087-V86 (1986)
and Lit. Fiz. Sb. {\bf 27}, 368 (1987) [Sov Phys.-Collect. {\bf 27}, 111
(1987)].

\item  V. Gontis and B. Kaulakys, J. Phys. B: At. Mol. Opt. Phys. {\bf 20},
5051 (1987).

\item  B. Kaulakys and G. Vilutis, in {\it Chaos - The Interplay between
Stochastic and Deterministic Behaviour,} edited by P. Garbaczewski, M. Wolf,
and A. Weron, Lecture Notes in Physics Vol. {\bf 457} (Springer-Verlag,
Berlin, 1995), p. 445; chao-dyn/9503011; B. Kaulakys, D. Grauzhinis and G.
Vilutis, Europhys. Lett. {\bf 43}, 123 (1998); physics/9808048.

\item  A. J. Lichtenberg and M. A. Lieberman, {\it Regular and Stochastic
Motion} ( Springer-Verlag, New York, 1983 and 1992).

\item  G. M. Zaslavskii, {\it Stochastic Behavior of Dynamical Systems}
(Nauka, Moscow, 1984; Harwood, New York, 1985).

\item  R. V. Jensen, Am. Scient. {\bf 75}, 168 (1987).

\item  B. V. Chirikov, Phys. Reports {\bf 52}, 265 (1979).

\item  S. Fahy and D. R. Hamann, Phys. Rev. Lett. {\bf 69}, 761 (1992).

\item  D. Park, {\it Classical Dynamics and Its Quantum Analogues}
(Springer, Berlin, 1990).

\item  L. E. Reichl, {\it The Transition to Chaos: In Conservative Classical
Systems: Quantum Manifestations} (Springer, Berlin, 1992).

\item  J. D. Meiss, Rev. Mod. Phys. {\bf 64}, 795 (1992).

\item  J. M. Ottino {\it et al.}, Science {\bf 257}, 754 (1992).

\item  B. Kaulakys and G. Vektaris, Phys. Rev. A {\bf 52}, 2091 (1995);
chao-dyn/9504009.

\item  F. Ivanauskas, T. Me\v skauskas and B. Kaulakys, {\it New Trends in
Prob. and Stat.}, Vol. 4, Eds. A. Laurin\v cikas {\it et al.} (TEV, Vilnius,
1997), p. 467.

\item  B. Kaulakys, F. Ivanauskas and T. Ma\v skauskas, {\it Proc. Intern.
Conf. on Nonlinearity, Bifurcation and Chaos: the Doors to the Future},
Lodz-Dobiesckow, Sept. 16-18, 1996; chao-dyn/9610020; Intern. J. Bifurcation
and Chaos (to be published).

\item  Y.-Y. Chen, Phys. Rev. Lett. {\bf 77}, 4318 (1996).

\item  B. Kaulakys and V. Gontis, Phys. Rev. A {\bf 56}, 1131 (1997);
quant-ph/9708024.

\item  R. J. Cook, Phys. Scr. {\bf T21}, 49 (1988).

\item  W. M. Itano, D. J. Heinzen, J. J. Bollinger, and D. J. Wineland,
Phys. Rev. A {\bf 41}, 2295 (1990).

\item  V. G. Gontis and B. P. Kaulakys, Liet. Fiz. Rink. {\bf 28}, 671
(1988) [Sov. Phys.-Collec. {\bf 28}(6), 1 (1988)].
\end{enumerate}

\end{document}